\documentclass[12pt]{article}
\usepackage{fullpage,amsmath}
\usepackage{setspace, lscape}
\usepackage{graphicx, psfrag, amsfonts, bm, amssymb}
\usepackage{wrapfig}
\usepackage{multirow}
\usepackage{hhline}
\usepackage{caption}
\usepackage{url, xr}
\usepackage[table]{xcolor}   
\usepackage{enumitem}

\usepackage[margin=1in]{geometry}

\makeatletter
\renewcommand\@biblabel[1]{}
\renewenvironment{thebibliography}[1]
     {\section*{\refname}%
      \@mkboth{\MakeUppercase\refname}{\MakeUppercase\refname}%
      \list{}%
           {\leftmargin0pt
            \@openbib@code
            \usecounter{enumiv}}%
      \sloppy
      \clubpenalty4000
      \@clubpenalty \clubpenalty
      \widowpenalty4000%
      \sfcode`\.\@m}
     {\def\@noitemerr
       {\@latex@warning{Empty `thebibliography' environment}}%
      \endlist}
\makeatother

\begin{document}
\begin{center}
{\large \bf  DROID: Dose-ranging Approach to Optimizing Dose in Oncology Drug Development}
\end{center}

\vspace{2mm}
\begin{center}
Beibei Guo$^1$ and Ying Yuan$^{2, *}$

{\em
$^1$Department of Experimental Statistics, Louisiana State University, Baton Rouge, LA 70803, USA\\
$^2$Department of Biostatistics, The University of Texas MD Anderson Cancer Center\\
Houston, TX 77030, USA \\
$^*$ Email: yyuan@mdanderson.org
}
\end{center}

\vskip 0.1in\noindent
\begin{center}
\textbf{Abstract}
\end{center}

\vspace{2mm}
In the era of targeted therapy, there has been increasing concern about the development of oncology drugs based on the ``more is better" paradigm, developed decades ago for chemotherapy. Recently, the US Food and Drug Administration (FDA) initiated Project Optimus to reform the dose optimization and dose selection paradigm in oncology drug development. To accommodate this paradigm shifting, we propose a \underline{d}ose-\underline{r}anging approach to \underline{o}ptim\underline{i}zing \underline{d}ose (DROID) for oncology trials with targeted drugs. DROID leverages the well-established dose-ranging study framework, which has been routinely used to develop non-oncology drugs for decades, and bridges it with established oncology dose-finding designs to optimize the dose of oncology drugs. DROID consists of two seamlessly connected stages. In the first stage, patients are sequentially enrolled and adaptively assigned to investigational doses to establish the therapeutic dose range (TDR), defined as the range of doses with acceptable toxicity and efficacy profiles, and the recommended phase 2 dose set (RP2S). In the second stage, patients are randomized to the doses in RP2S to assess the dose-response relationship and identify the optimal dose. The simulation study shows that DROID substantially outperforms the conventional approach, providing a new paradigm to efficiently optimize the dose of targeted oncology drugs.

\vspace{5mm}

\section{Introduction}
\label{s:intro}

The recent approval of sotorasib by the US Food and Drug Administration (FDA) for metastatic non-small-cell lung cancers harboring the KRAS G12C mutation is groundbreaking. It marks the first approved drug to target KRAS, which had been considered ``undruggable" for decades. Despite the success, the development of sotorasib was hindered by a lack of robust dose exploration, which led the FDA to require Amgen to conduct a postmarketing trial to compare the approved 960-mg dose with a 240-mg dose. The FDA has required that Amgen complete the dose optimization study by October 2022 and submit a final study report by February 2023, also indicating in the multi-discipline review that this postmarketing study will inform possible future labeling updates.

One cause of the requirement to do the postmarketing trial is that sotorasib was developed based on the conventional more-is-better paradigm. The phase I dose escalation trial was conducted using the Bayesian logistic regression model (Neuenschwander, Branson and Gsponer, 2008), a variation of the continuous reassessment method (CRM; O'Quigley, Pepe and Fisher, 1990), with the goal of finding the maximum tolerated dose (MTD) that yields a dose limiting toxicity (DLT) probability in the range of 0.20 to 0.30. Per the design, the highest dose of 960 mg was selected as the MTD, although 58 patients (45\%) were classified as having serious complications and nine patients (7\%) discontinued treatment because of toxicity. This MTD was subsequently used for the phase II registration trial, leading to the approved dose.

	There has been increasing concern regarding the appropriateness of using the more-is-better paradigm in the era of targeted therapies and immunotherapies (Ratain, 2014; Zang, Lee and Yuan, 2014; Yan, Thall, and Yuan, 2018; Ratain et al., 2021; Shah et al., 2021). Unlike cytotoxic agents, most novel targeted agents have a different mechanism of action and work by inhibiting molecular pathways that stimulate proliferation of cancer cells or inhibit their death.  With targeted drugs, increasing doses beyond a certain level may not enhance antitumor activity. In addition, DLT effects may not be observed at clinically active doses. Therefore, for these agents, dosing at the MTD is often inappropriate, leading to off-target effects including toxicity, dose interruptions, and reduced compliance. The objective of early phase trials for targeted drugs should be to establish the optimal dose, defined as the dose that is safe and above which there is no evidence of incremental benefit. The requirement of a postmarketing study for sotorasib signals the regulatory agency's increasing emphasis on optimizing the risk-benefit for patients and new scrutiny around determining the labeled dose for targeted agents. Recently, FDA Oncology Center of Excellence initiated Project Optimus ``to reform the dose optimization and dose selection paradigm in oncology drug development" (FDA, 2022). 

	To accommodate this paradigm shift, we propose a \underline{d}ose-\underline{r}anging approach to \underline{o}ptim\underline{i}zing \underline{d}ose (DROID) for oncology clinical trials with targeted agents. DROID bridges the dose-ranging study framework, routinely used in the development of non-oncology drugs for decades, with the established oncology dose-finding designs to provide a robust and efficient method to optimize the dose for oncology. DROID consists of two seamlessly connected stages. In the first stage, patients are sequentially enrolled and adaptively assigned to investigational doses to find the therapeutic dose range (TDR), defined as the range of doses with acceptable toxicity and efficacy profiles, and recommended phase 2 dose set (RP2S).  In the second stage, patients are randomized to RP2S to assess the dose-response relationship and identify the optimal dose. 
	
	Along a different line, a number of phase I-II designs have been proposed to find the optimal dose that maximizes a certain risk-benefit criterion. Braun (2002) generalized CRM to accommodate toxicity and efficacy simultaneously. Thall and Cook (2004) presented the EffTox design, based on the trade-offs between toxicity and efficacy. Yin et al (2006) proposed a phase I-II design that uses the odds ratio of the efficacy and toxicity as a measure of desirability. Yuan and Yin (2009) described a phase I-II design that jointly models toxicity and efficacy as time-to-event outcomes. Jin et al. (2014) proposed a phase I-II design that accommodates late-onset toxicity and efficacy. Liu and Johnson (2016) proposed a robust Bayesian phase I-II design, based on a flexible Bayesian dynamic model. Guo and Yuan (2017) proposed a personalized Bayesian phase I-II design that accounts for patient characteristics and biomarker information. Yuan et al (2016) provided a comprehensive coverage of phase I-II designs. These model-based designs are statistically complicated and require real-time model estimation to make sequential decisions of dose assignment, limiting their use in practice. Some recent efforts have been dedicated to developing model-assisted designs, which derive and optimize their decision rules based on models, but can be implemented in a similar way as algorithm-based designs (e.g., 3+3 design) by simply looking up the decision table (Yuan et al., 2019). Model-assisted phase I-II designs include utility-based Bayesian optimal interval design (U-BOIN) (Zhou, Lee, and Yuan, 2019), BOIN with efficacy and toxicity endpoints (BOIN-ET) (Takeda, Taguri, and Morita, 2018), and the BOIN12 design (Lin et al., 2020). These designs are simple to implement and deliver comparable performance to more complicated model-based designs. All aforementioned designs were developed within the oncology dose-escalation framework, despite incorporating the risk-benefit consideration, with limited consideration of identification of TDR/RP2S, pharmacodynamics (PD) endpoint, randomization within RP2S, or assessment of the dose-response relationship.

	The remainder of this paper is organized as follows. In Section 2, we propose DROID designs. In Section 3, we present simulation studies to show the operating characteristics of DROID designs compared with the conventional design. We conclude with a brief discussion in Section 4.

\section{Methods}
\label{s:model}

\subsection{Stage I of Finding the TDR}

Consider a trial with $J$ doses under investigation, $d_1<d_2<\cdots<d_J$. Let $Y_T$ denote the binary toxicity endpoint (with $Y_T=1$ indicating toxicity), and $Y_E$ denote a binary efficacy endpoint such as tumor response (with $Y_E=1$ indicating response). Let $Y_S$ denote a continuous PD biomarker, measuring the biological antitumor activity of the drug, or any other surrogate endpoint for efficacy. For ease of exposition and without loss of generality, we assume that $Y_S$ is a PD biomarker. We consider a common scenario that $Y_T$ and $Y_S$ are quickly ascertainable (e.g., within the first cycle of treatment), whereas the evaluation of  $Y_E$ may take a substantially longer time (e.g., requiring multiple cycles), which hinders timely decision making. To address this logistic challenge, the DROID design proposes the gatekeeping approach, where adaptive decisions (e.g., dose assignment) are made based on $Y_T$ and $Y_S$, while $Y_E$ is used as a gatekeeper at some key decision-making points (e.g., when selecting RP2S for stage II randomization and when selecting the optimal dose at the end of the trial) to ensure that the doses selected have acceptable clinical efficacy.

Let $p(d_j)=\mbox{Pr}(Y_T=1|d_j)$ denote the toxicity probability, and $\mu (d_j)=E(Y_S |d_j)$ denote the mean PD measure of $d_j$. Let $\phi_T$ denote the highest acceptable toxicity rate, and $\phi_S$ denote the lowest acceptable PD threshold. The TDR is defined as the set of doses with  $p(d_j)\leq\phi_T$ and $\mu(d_j )\geq\phi_S$, i.e., 
$$\mbox{TDR} = \{d_j: p(d_j)\leq\phi_T\ \ \  \&\ \ \  \mu(d_j )\geq\phi_S \}.$$
Under the generally valid assumption that dose-toxicity and dose-PD curves are non-decreasing (including plateau), identification of the TDR boils down to pinpointing the minimal active dose (MAD), defined as the lowest dose with $\mu(d_j )\geq\phi_S$ (the lower bound of the TDR), and the MTD, defined as the highest dose with $p(d_j) \leq\phi_T$ (the upper bound of the TDR). In other words, the TDR is the range of doses between the MAD and MTD. Here, the MAD is analogous to the minimal effective dose that is routinely used in non-oncology dose-finding studies (Zhou, et al., 2017). In what follows, we consider two strategies to find the TDR: model-based approach and model-assisted approach.

\subsubsection{Model-based approach}
\noindent\textit{Dose-Toxicity and -PD Models}\\
The model-based approach uses parametric models to describe dose-toxicity and dose-PD relationships and guide dose transition and identification of TDR. Assuming at an interim decision time, $n_j$ patients have been treated at $d_j$. Let $Y_{T,ji}$ and $Y_{S,ji}$ denote the toxicity and PD endpoints of the $i$th patient at $d_j$, where $i=1,\cdots,n_j$ and $j=1,\cdots,J$. Let $\tilde{d}_j$ denote the effective dose as described below. We model the dose-toxicity curve using a logistic regression model,
\begin{equation}
\mbox{logit}\big(p(d_j) \big)=\alpha_0+\alpha \tilde{d}_j
\end{equation}
where $\alpha_0$ and $\alpha$ are model parameters. Following O'Quigley and Chevret (1991) and Chevret (1993), we fix $\alpha_0=3$. Let $q_j$ be the prior estimate of the toxicity probability at dose level $j$ (a.k.a., skeleton), and   $\hat{\alpha}$ denote the prior estimate of $\alpha$ (e.g., the mean of the prior distribution). Then the effective dose $\tilde{d}_j$ is determined by back-solving the dose-toxicity model $\tilde{d}_j=\big\{\mbox{log}(\frac{q_j}{1-q_j })-\alpha_0 \big\}/\hat{\alpha}$. Research shows that the resulting one-parameter logistic model yields more robust and better performance than the two-parameter logistic model that regards $\alpha_0$ as an unknown parameter (Chevret, 1993; Iasonos, et al., 2016). 

For PD endpoint $Y_S$, we employ a Bayesian Emax model
\begin{equation*}
Y_{S,ji}|d_j=\mu (d_j)+\epsilon_{ji}, \ \ \ \epsilon_{ji}\stackrel{iid}{\sim} N(0,\sigma^2)
\end{equation*}
\begin{equation}
\mu(d_j)=\eta+\frac{\tau d_j^{\gamma}}{\beta^{\gamma}+d_j^{\gamma}}
\end{equation}
where $\eta$ is the baseline activity value without the drug; $\tau$ is the maximum activity that is possibly achieved (i.e., Emax) with the drug above the baseline activity, indicating where the dose-PD curve plateaus; $\beta$ is the dose that produces half of the Emax (i.e., ED50); and $\gamma$ is the Hill factor that controls the steepness of the dose-response curve. The Emax model is chosen because it accommodates the important feature that $Y_S$ may plateau after a certain $d_j$, and also provides a good fit to real world data based on a large scale empirical study (FDA, 2022). Nevertheless, other dose-response models (e.g., quadratic model) or a parsimonious Emax model without the parameter of the Hill factor can be entertained when appropriate. We choose to model the marginal distributions of $Y_T$ and $Y_S$, rather than their joint distribution, because the identification of TDR only involves the marginal distributions. This approach simplifies the model estimation with little impact on the operating characteristic of the design as shown by the simulation study later. 

Given the interim data $D=\{y_{T,ji}, y_{S,ji}\}$, the likelihood for toxicity is
\begin{equation*}
L_T (D|\alpha)=\prod_{j=1}^J\prod_{i=1}^{n_j}\Big(\frac{\mbox{exp}(\alpha_0+\alpha \tilde{d}_j)}{1+\mbox{exp}(\alpha_0+\alpha \tilde{d}_j)} \Big)^{y_{T,ji}} \Big(\frac{1}{1+\mbox{exp}(\alpha_0+\alpha \tilde{d}_j)} \Big)^{1-y_{T,ji}},
\end{equation*}
and the likelihood for PD is
\begin{equation*}
L_S (D|\eta,\beta,\tau,\gamma)=\prod_{j=1}^J\prod_{i=1}^{n_j}\frac{1}{\sigma\sqrt{2\pi}}  \mbox{exp}\Bigg\{ -\frac{1}{2} \bigg(\frac{y_{S,ji}-\eta-\frac{\tau d_j^{\gamma}}{\beta^{\gamma}+d_j^{\gamma}}}{\sigma} \bigg)^2 \Bigg\}.
\end{equation*}
Let $f(\alpha)$ and $f(\eta,\beta,\tau,\gamma)$ denote the prior distributions, the posterior distributions are given by
\begin{equation*}
f(\alpha|D)\propto f(\alpha) L_T (D|\alpha),
\end{equation*}
\begin{equation*}
f(\eta,\beta,\tau,\gamma|D)\propto f(\eta,\beta,\tau,\gamma) L_S (D|\eta,\beta,\tau,\gamma),
\end{equation*}
which can be sampled using the Gibbs sampler. Guidance on prior specification is provided in the Supplementary Materials.

\noindent\textit{TDR-Finding Algorithm}\\
We find TDR by employing two parallel dose exploration processes: the MAD-finding process and MTD-finding process. In what follows, we first describe the decision rules of the two processes, followed by the TDR-finding algorithm. 

\noindent (i) MAD-finding decision rule:\\
Let $j_S$ denote the current dose level for the MAD-finding process, and $j_S^*$ denote the dose level whose posterior estimate of $\mu(d_j)$ is closest to $\phi_S$, i.e., $j_S^*=\underset{j\in\{1,\cdots,d\}}{\arg\min}|\hat{\mu}(d_j)-\phi_S |$
\begin{itemize}
\item If $j_S^*>j_S$, the candidate dose for the next cohort is $\tilde{d}_S =\mbox{min}(j_S+1,J)$.
\item If $j_S^*<j_S$, the candidate dose for the next cohort is $\tilde{d}_S =\mbox{max}(j_S-1,1)$.
\item If $j_S^*=j_S$, the recommended dose for the next cohort is $\tilde{d}_S=j_S$.
\end{itemize}

\noindent (ii) MTD-finding decision rule:\\
Let $j_T$ denote the current dose for the MTD-finding process, and $j_T^*$ denote the dose level whose posterior estimate of $p(d_j)$ is closest to $\phi_T$, i.e., $j_T^*=\underset{j\in\{1,\cdots,d\}}{\arg\min}|\hat{p}(d_j)-\phi_T  |$.
\begin{itemize}
\item If $j_T^*>j_T$, the recommended dose for the next cohort is $\tilde{d}_T =\mbox{min}(j_T+1,J)$.
\item If $j_T^*<j_T$, the recommended dose for the next cohort is $\tilde{d}_T =\mbox{max}(j_T-1,1)$.
\item If $j_T^*=j_T$, the recommended dose for the next cohort is $\tilde{d}_T=j_T$.
\end{itemize}
In (i) and (ii), we do not allow dose skipping. For (i), de-escalating the dose when $j_S^*<j_S$ is desirable for reliable identification of TDR/RP2S, which will be later used to identify the optimal dose, for the simple reason that we have to concentrate data on the part of the dose-response curve of interest to obtain reasonable power. This efficient use and allocation of data is particularly important here because of the small sample size of stage I. 

With the above decision rules at hand, the TDR-finding algorithm is described as follows:
\begin{enumerate}
\item 	Treat the first cohort of patients at the lowest dose with $j_S=j_T=1$. 
\item 	Given the observed data, apply MAD-finding and MTD-finding decision rules to determine the candidate dose levels $\tilde{d}_T$ and $\tilde{d}_S$ for the next cohort: 
\begin{itemize}
\item 	If $\tilde{d}_T\leq\tilde{d}_S$, treat the next cohort of patients at $\tilde{d}_T$.
\item 	Otherwise, randomize patients between $\tilde{d}_T$ and $\tilde{d}_S$, such that each of the doses receives a cohort of patients.
\end{itemize}
\item 	Repeat Step 2 until reaching a prespecified maximum sample size $N_1$ for stage I, or early stop the trial if $\mbox{Pr}(p(d_1)>\phi_T|\mbox{data})>C_{T1}$ (toxicity stopping when the lowest dose is overly toxic) or  $\mbox{Pr}(\mu(d_J)<\phi_S|\mbox{data})>C_{S1}$ (PD stopping when the highest dose is futile), where $C_{T1}$ and $C_{S1}$ are probability cutoffs. 
\end{enumerate}

As demonstrated in Step 2, one feature of DROID is that the two dose-finding processes merge or split adaptively according to the observed data, making it different from existing dose-finding designs. If MTD is coincident with MAD, the two processes automatically merge as one, improving the efficiency of the design. For example, when the low doses are safe but subtherapeutic, the two processes will merge as one, leading to fast dose escalation over the low doses and allocating more patients to effective doses. On the other hand, when MTD differs from and is higher than MAD, the design splits the two processes and assigns patients concurrently to each of them. This is sensible because the correct identification of TDR hinges on the correct identification of both MAD and MTD. Concentrating patients around these two target doses renders the design more capable of learning and identifying TDR with greater reliability and efficiency. This is analogous to backfilling patient during the dose escalation, an empirical approach commonly used in practice,  but the proposed approach is  more principled and rigorous with explicit statistical decision criteria and objectives. The proposed TDR-finding design is a dual-target finding design, whereas most existing designs are single-target finding designs.

When stage I is complete, we identify MTD as the highest dose whose $\hat{p}(d_j) \leq\phi_T$, and MAD as the lowest dose whose $\hat{\mu}(d_j)\geq\phi_S$. The TDR are the doses between MAD and MTD, i.e., TDR = $\{d_j,\mbox{MAD}\leq d_j \leq \mbox{MTD}\}$. An alternative approach is to select TDR from the continuous dose range $[d_1,d_J]$ as TDR = $[\mbox{max}(d_1,F_S^{-1} (\phi_S)), \mbox{min}(d_J,F_T^{-1} (\phi_T))]$, where $F_S^{-1}(.)$ and $F_T^{-1}(.)$ denote the inverses of dose-PD and dose-toxicity functions (1) and (2). If dose extrapolation is allowed beyond the investigational dose range, TDR is given by $[ F_S^{-1} (\phi_S), F_T^{-1}(\phi_T)]$. Dose extrapolation below $d_1$ might be acceptable, but extrapolation to doses higher than $d_J$ is often problematic due to safety concerns. If TDR is empty, the trial is terminated with all doses considered unacceptable. 

TDR forms the basis of determining the recommended phase 2 dose set (RP2S) --- the doses to be moved forward to stage II for randomization. Let $\pi(d_j )=\mbox{Pr}(Y_E=1|d_j)$ denote the objective response rate (ORR) of $d_j$, and $\phi_E$ denote the lowest acceptable ORR. We define RP2S as the subset of TDR that satisfies the efficacy requirement $\hat{\pi}(d_j)>\phi_E$, where $\hat{\pi}(d_j)$ is the posterior mean estimate of $\pi(d_j)$, obtained by applying beta-binomial model on $Y_E$ (see the Supplementary Materials). Alternatively, the posterior-probability-based efficacy requirement, e.g., $\mbox{Pr}(\pi(d_j)>\phi_E |data)>C_E$, can also be used. The efficacy requirement is used to gatekeep the case that a dose showing sufficient PD effect may not always demonstrate sufficient clinical efficacy. As described previously, we do not directly use $Y_E$ to guide dose escalation/de-escalation because it often takes a long time to be ascertained.  In addition to $Y_E$, dose tolerability, measured by dose interruption and reduction over multiple cycles, can also be used to guide the choice of RP2S. Dose tolerability by definition typically requires a long time to be observed, thus is more suitable as a gatekeeping endpoint, rather than guiding dose escalation/de-escalation. We assume that there is a follow-up period (e.g., 2-3 months) after stage I enrollment so that $Y_E$ and dose tolerability are observed for all or most stage I patients.  To reduce the sample size and cost, in practice, it is often desirable to impose a constraint that the number of doses in RP2S cannot be more than $K$ (e.g., 3) by choosing the $K$ doses with the highest $\hat{\pi}(d_j)$ within TDR as RP2S.

\subsubsection{Model-assisted approach}
This section briefly discusses the model-assisted approach to determine TDR. The advantage of this approach is its simplicity, which does not require complicated model fitting to make the decision of dose transition, and competitive performance comparable to model-based designs (Zhou et al., 2018). A review of model-assisted designs is provided by Yuan et al. (2019, 2022). We here adopt the BOIN design as the basis to guide the MAD-finding and MTD-finding processes. Let $\hat{p}_j=\sum_{i=1}^{n_j}y_{T,ji}/{n_j}$  denote the observed toxicity rate and $\hat{\mu}_j=\sum_{i=1}^{n_j}y_{S,ji}/{n_j}$ denote the sample mean of $Y_S$ at $d_j$. Similarly as the model-based approach, TDR finding is based on the MAD-finding and MTD-finding processes.

\textit{MAD-finding decision rule}:\\
Let $j_S$ denote the current dose level for the MAD-finding process, and $\gamma_e$ and $\gamma_d$ denote prespecified dose escalation and de-escalation boundaries for PD, respectively.
\begin{itemize}
\item If $\hat{\mu}_{j_S}\leq\gamma_e$, the candidate dose for the next cohort is $\tilde{d}_S=\mbox{min}(j_S+1,J)$.
\item If $\hat{\mu}_{j_S}>\gamma_d$, the candidate dose for the next cohort is $\tilde{d}_S=\mbox{max}(j_S-1,1)$.
\item Otherwise, the candidate dose for the next cohort is $\tilde{d}_S=j_S$.
\end{itemize}

\textit{MTD-finding decisioin rule}:\\
Let $j_T$ denote the current dose for the MTD-finding process, and $\lambda_e$ and $\lambda_d$ denote prespecified dose escalation and de-escalation boundaries for toxicity, respectively.
\begin{itemize}
\item If $\hat{p}_{j_T}\leq\lambda_e$, the candidate dose for the next cohort is $\tilde{d}_T=\mbox{min}(j_T+1,J)$.
\item If $\hat{p}_{j_T}>\lambda_d$, the candidate dose for the next cohort is $\tilde{d}_T=\mbox{max}(j_T-1,1)$.
\item Otherwise, the candidate dose for the next cohort is $\tilde{d}_T=j_T$.
\end{itemize}

Liu and Yuan (2015) provided default optimal values of $\lambda_e$ and $\lambda_d$ for common DLT rates. For example, $(\lambda_e, \lambda_d)= (0.157, 0.238), (0.197, 0.298), (0.236, 0.358), (0.276, 0.419)$ for $\phi_T= 0.2, 0.25, 0.3, 0.35$, and 0.4, respectively. For continuous PD endpoint $Y_S$, Mu et al. (2019) recommended $\gamma_e=0.8\phi_S$ and $\gamma_d=1.2\phi_S$, which generally yield desirable operating characteristics. These boundaries can be further calibrated using simulation to fit specific trial considerations. Following BOIN, we impose the following overdosing and underdosing control rules: eliminate $d_j$ and higher doses for toxicity if $\mbox{Pr}(p(d_j)>\phi_T |data)>C_T$, where $C_T$ is a probability cutoff (e.g., $C_T=0.95$); and eliminate $d_j$ and lower doses for futility if $\mbox{Pr}(\mu(d_j)<\phi_S |data)>C_S$, where $C_S$ is a probability cutoff (e.g., $C_S=0.95$). If all doses are eliminated due to toxicity or futility, the trial should be terminated.

Given the above MTD- and MAD-finding decision rules, the TDR-finding algorithm is the same as that for the model-based approach. At the end of stage I, we identify TDR with MTD as the highest dose whose isotonic estimate (Robertson, Write, and Dykstra, 1988) of $p(d_j)\leq \phi_T$, and MAD as the lowest dose whose isotonic estimate of $\mu(d_j)\geq\phi_S$. When desirable, the TDR can also be selected in the same way as previously described by fitting dose-toxicity and -PD models. As the dose transition and TDR determination are independent tasks, by doing so, we use the model-assisted approach to facilitate the trial conduct, while also leveraging the model-based approach to obtain extra flexibility to select TDR (e.g., select TDR in the continuous dose range).

\subsection{Stage II of Randomization}
At stage II, we randomize more patients to the doses within RP2S to assess the dose-response relationship. Depending on the trial objectives and characteristics, different randomization schemes can be used. The most straightforward approach is to fix the sample size of stage II and equally randomize patients to the doses in the identified RP2S. As the number of patients treated in stage I is often different across doses, equal randomization results in different total numbers of patients across the doses in RP2S. Thus, an alternative randomization scheme is to choose a randomization ratio, based on the number of patients at stage I, such that at the end of stage II, an equal total number of patients will be treated at each RP2S dose. A more sophisticated randomization scheme is the outcome-dependent adaptive randomization. For example, we randomize patients to the doses within the RP2S with probabilities proportional to their desirability. The desirability can be defined as $\hat{\mu}(d_j)$ or the risk-benefit tradeoff using utility (Zhou, et al., 2019). 

Regardless of the randomization scheme employed, throughout stage II, we apply the following toxicity and futility rules to drop toxic or ineffective doses in a continuous or group-sequential fashion, as follows: 
\begin{itemize}
\item 	(Safety rule) Drop $d_j$ and higher doses for toxicity if $\mbox{Pr}(p(d_j)>\phi_T |data)>C_{T,2}$, where $C_{T,2}$ is a probability cutoff that is calibrated using simulation. 
\item 	(Futility rule) Drop $d_j$ and lower doses for futility if $\mbox{Pr}(\mu(d_j)<\phi_S |data)>C_{S,2}$, where $C_{S,2}$ is a probability cutoff that is calibrated using simulation.
\end{itemize}

For model-assisted DROID, $\mbox{Pr}(p(d_j)>\phi_T |data)$ and $\mbox{Pr}(\mu(d_j)<\phi_S |data)$ can be evaluated based on isotonic transformed posterior distributions of $p(d_j)$ and $\mu(d_j)$ across all tried dose levels or model estimates (e.g., logistic model (1) for toxicity and Emax model (2) for PD) as in the model-based approach. During stage II, when desirable, at each interim, TDR/RP2S could be updated based on accumulating data, and new doses that become quantified as TDR/RP2S doses in light of new data may be added to stage II for randomization. As shown later in the simulation study, this may improve the performance in some scenarios, but is more involved logistically and operationally in practice.

Stage II randomization is an analog of dose-ranging studies that are routinely used, and often required by the FDA in the development of non-oncology drugs. In dose ranging studies, patients are randomized to several dose groups, often with a placebo included as a control. The objective of the dose ranging study is to achieve (i) proof of concept (PoC) by establishing the statistical significance of the dose-response relationship, and (ii) identify the minimal effective dose, defined as the lowest dose that achieves a prespecified clinical relevance (in terms of the point estimate) and also statistical significance compared to the control. A variety of designs, e.g., multiple comparison procedures with modeling techniques (MCP-Mod), are developed to achieve these two objectives, see Dragalin et al. (2010) for a review on the dose-ranging methods.

Dose ranging studies, however, are rarely used in oncology for various reasons. First, objective (i) of dose ranging studies is evaluated based on the classic hypothesis testing framework with multiplicity adjustment to control familywise type I error (e.g., MCP-Mod). This approach demands a large sample size (e.g., hundreds of patients), which often is not feasible in early phase oncology trials due to the difficulty of accrual and tight development timeline. Second, objective (ii) of dose ranging studies often include a placebo group, which is often not feasible in oncology either. Third, as the demonstration of efficacy tends to be the biggest hurdle associated with oncology drug approval, sponsors often lean toward selecting the dose with the highest efficacy, rather than the minimal dose that reaches the clinical relevance. Despite the differences and challenges, the general principle and approach of dose-ranging studies are valuable to improve the development of targeted oncology drugs, as described next.

\subsection{Assessment of dose-response relationship and identification of the optimal dose}
To leverage the decades of experience and success of non-oncology dose-ranging studies, we propose a new oncology dose-ranging approach with two objectives that parallel with those of dose-ranging studies, while accounting for the unique characteristics and challenges of oncology trials. The estimation and decisions are based on the overall data from stages I and II.

The first objective is to establish PoC. We proposed to use the Bayesian dose-response index (DRI), rather than the p-value and hypothesis testing, to assess the strength of evidence on the dose-response relationship. We define DRI as:
\begin{equation}
 \mbox{DRI}=\mbox{Pr}(\mu(d_1 )<\delta\mu (d_{H^*})|data)                                       
\end{equation}
where $d_{H^*}$ is the highest tried dose, and $\delta$, say $\delta=0.9$, is the equivalence margin (i.e., PD is regarded as plateau at a dose if its PD level is greater than $\delta$ times that of the highest tried dose). DRI represents the probability of the existence of a dose-response relationship given the observed data, providing a direct measure on the evidence of the dose-response relationship. For example, DRI = 0.8 means that there is an 80\% chance that a dose-response relationship exits, or the odds of existence versus absence of a dose-response relationship is 4:1. The DRI in (3) assumes a non-decreasing dose-PD relationship, which is plausible in most cases. If an umbrella-shape relationship is perceived, we can replace  $d_{H^*}$ with the dose with the highest estimate of $\mu(d_j)$. Let $C_{DRI}$ be a probability cutoff to be tuned through simulation. If $\mbox{DRI} > C_{DRI}$, we claim the dose-response relationship established. This intuitive interpretation facilitates the sponsor and regulatory agent to evaluate and determine whether a specific value of DRI is sufficient to establish PoC based on the characteristics of the drug and targeted patient population, rather than universally applying the 0.05 cutoff for the p-value and ignoring the different characteristics of each drug development. The flaws of using a p-value and significance as the metric of evidence have been extensively discussed in the literature (Johnson, 2013). In the case of observing strong evidence of no dose-response relationship, i.e., when $\mbox{DRI} < C_{DRI}$, a follow-up randomization study with additional low doses may be conducted to further characterize the dose-response relationship when appropriate. Another advantage of DRI is that it is straightforward for it to account for the uncertainty of the dose-response model using Bayesian model averaging (BMA) to improve the robustness of the inference along a similar line as MCP-Mod when appropriate (see the Supplementary Materials).

The second objective is to select the optimal dose. This differs from that of non-oncology dose ranging studies, which aims to identify the minimal effect dose. This modification is necessary as the demonstration of efficacy tends to be the biggest hurdle associated with oncology drug approval. Let $S=\{d_L<\cdots<d_H\}$ denote the doses in RP2S that are not dropped due to toxicity or futility at the end of stage II. Given that PoC has been established, we select the optimal dose $d_{opt}$ as the lowest dose that reaches the PD plateau and clinical relevance. 
\begin{equation}
d_{opt}=min\{d_j: \mbox{Pr}(\mu(d_j )\geq \delta\mu(d_H)|data)>C_1 \ \  \& \ \ \mbox{Pr}(\pi(d_j)\geq\phi_E |data)>C_2,d_j\in S\}      
\end{equation}
where $C_1$ and $C_2$ are probability cutoffs. Other statistical criteria can also be used to define $d_{opt}$, for example based on the point estimates.
\begin{equation}
d_{opt}=\mbox{min}\{d_j: \hat{\mu}(d_j )\geq\delta\hat{\mu}(d_H )\ \  \&\ \   \hat{\pi}(d_j)\geq\phi_E,d_j\in S\}.                     
\end{equation}
In the case that PoC cannot be established (i.e., the dose-response relationship is fairly flat), additional higher or lower doses outside $[d_1,d_J]$ may be further studied, or the optimal dose may be selected based on the totality of data and clinical considerations. In our simulation, to simplify reporting results, when PoC cannot be established, we do not select the optimal dose. The optimal dose selection criteria (4) and (5) assume that when PD plateaus, the efficacy also plateaus, which is often reasonable for targeted drugs. We investigated an alternative approach of selecting $d_{opt}$ directly based on ORR as the dose where $\pi(d_j)$ plateaus. This approach does not require the above assumption, but does not work well as the binary tumor response endpoint provides very limited power to identify the plateau point (see the Supplementary Materials).

\section{Simulation Study}
We evaluated the operating characteristics of the DROID design using simulation studies. We considered five doses (0.1, 0.3, 0.5, 0.7, 0.9) and assumed the prior estimate of toxicity probabilities (0.05, 0.15, 0.3, 0.4, 0.55) to obtain the effective doses $\tilde{d}_j$. The stage I sample size was 12 cohorts of size 3. In stage II, the maximum sample size for each dose is M = 20. The toxicity upper bound $\phi_T=0.3$, the PD lower bound $\phi_S=0.1$, and the ORR lower bound $\phi_E=0.3$. For $\eta,\beta$, and $\tau$, the clinician-elicited prior estimates were 0.1, 0.5, and 0.4, respectively with elicited ranges (0, 0.3), (0, 1), and (0.1, 0.7). This resulted in priors $\eta\sim\mbox{Gamma}(1, 10)$,   $\beta\sim\mbox{Gamma}(4, 8)$, and $\tau\sim\mbox{Gamma}(7.1, 17.8)$ based on the prior specification procedure described in the Supplementary Materials. For the Hill factor $\gamma$, the elicited prior estimate was 2, so we set $\gamma\sim\mbox{Gamma}(1/9, 1/18)$ such that the prior standard deviation was three times the prior mean. The equivalence margin was taken as $\delta$=0.9. Calibrated by simulation, we took the probability cutoffs $C_T=C_E=C_{T,2}=C_{S,2}=0.95$, $C_{DRI}=0.7,C_1=0.37$, and $C_2=0.15$. 

We considered nine scenarios that varied in the location of the optimal dose and the patterns of toxicity, PD, and ORR (see Table 1). Figure 1 shows the true dose-response curves for toxicity, PD, and ORR for these scenarios. Under each scenario, we simulated 1,000 trials.
To simulate correlated toxicity, PD, and ORR data for patient $i$, we first generated the PD data $Y_{S,i}$  from the Emax model. Conditional on $Y_{S,i}$, we generated toxicity and ORR by  
\begin{equation*}
\mbox{Logit}\big(\mbox{Pr}(Y_{T,i})=1 \big)=\xi_0+\xi_1Y_{S,i}+\theta_i
\end{equation*}
\begin{equation*}
\mbox{Logit}\big(\mbox{Pr}(Y_{E,i})=1 \big)=\zeta_0+\zeta_1Y_{S,i}+\theta_i
\end{equation*}
where $\theta_i\sim N(0,\tau_0^2)$, $\tau_0^2=1$, is a patient-specific random effect used to induce positive correlation between $Y_T$ and $Y_E$. We compared the DROID design with the EffTox design (Thall and Cook, 2006), which aims to find the optimal dose based on efficacy-toxicity tradeoff, and CRM. We chose EffTox as a comparator because it is one of the a few dose-optimization designs that have been implemented in real trials (deLima et al., 2008; Konopleva et al, 2015; Tidwell et al., 2021; Msaouel et al., 2022). We included CRM mainly to demonstrate the difference between finding the MTD and optimal dose, while noting that such comparison may not be fair given that the two designs have different objectives. We let DROID-CRM and DROID-BOIN denote the two versions of the DROID design that use CRM or BOIN in stage I, respectively.

Table 2 summarizes the operating characteristics of the DROID designs, EffTox, and CRM. In scenario 1, the first four dose levels are safe and PD plateaus from dose level 2, so the optimal dose is dose level 2 with ORR = 0.5. The percentage of correct selection (PCS) of the optimal dose under DROID-CRM and DROID-BOIN are over 75\%, that of CRM and EffTox is 5.1\% and 35\%, respectively. In scenarios 2 and 3, the optimal dose is dose level 3 as it is safe and PD plateaus from dose level 3. Like in scenario 1, the PCS of DROID designs is over 76\%, substantially higher than those of CRM and EffTox. For scenarios 4 and 5, the optimal dose is dose level 4. Like for scenarios 2 and 3, the DROID designs yielded higher PCS than CRM and EffTox. In scenario 6, PD keeps increasing with dose. Since dose level 3 is overly toxic, dose level 2 is the optimal dose. In scenario 7, PD increases with dose, so the optimal dose is dose level 5. In scenario 8, dose level 1 is the optimal dose as it is the only acceptable dose in terms of toxicity. In these three scenarios, DROID designs and CRM yielded similar PCS. In scenario 10, the PD curve is almost flat. The DROID designs claimed no PoC about 99\% of the time while CRM and EffTox recommended a dose for each simulated trial by design.

We further investigated operating characteristics of the DROID designs when (i) the doses that are not included in TDR/RP2S at the end of stage I can be added during stage II randomization in light of accumulative data; and (ii) the alternative point-estimate-based criterion (5) is used to select the optimal dose at the end of the trial. Table 3 shows the results for (i). Although in some scenarios (e.g., scenarios 2 and 7), there are sizeable performance improvements because of the use of this additional adaptation rule, the performance is comparable in most scenarios. Therefore, if adding new doses during randomization is operationally challenging, the adaptation rule (i) may not be needed. Figures 2 and 3 provide the results for (ii), showing that selection percentages are similar to those when posterior distributions were used to identify the optimal dose.

\section{Discussion}
We have proposed DROID designs, a dose-ranging approach to optimizing dose for oncology trials with targeted drugs. DROID is a two-stage design. In the first stage, a model-based or model-assisted design is used to identify the TDR and RP2S in terms of both efficacy and toxicity. In the second stage, more patients are randomized to the doses in the TDR to assess the dose-response relationship for PoC and identify the optimal dose. Our simulation study shows that the proposed DROID designs have desirable operating characteristics.

To facilitate the implementation of the DROID design in practice, we use the PD (or efficacy surrogate endpoint) $Y_S$, along with the toxicity endpoint $Y_T$, to make the decision of dose assignment in stage I. In some trials, $Y_S$ may not be a strong surrogate for efficacy $Y_E$. This, however, is not of concern because $Y_S$ is mainly used to determine the relative desirability among doses, not to estimate efficacy of the doses. Thus, as long as $Y_S$ and $Y_E$ are reasonably concordant, patient allocation based on $Y_S$ is generally similar to that based on $Y_E$. Moreover and more important, in the DROID design,  $Y_E$ is indeed used as a gatekeeper for key decisions, e.g., to identify RP2S at the end of stage I and to establish PoC and the optimal dose at the end of the trial. An alternative approach is to build a regression model and use $Y_S$ to predict $Y_E$, and then make dose assignment decisions based on $Y_E$ and $Y_T$. As $Y_E$ often takes a long time to be observed and the number of events (i.e., responses) may be small, this approach may be of limited value and also subject to the influence of misspecification of the prediction model. 

Lastly, the stages I and II of the DROID design are not binding. In some applications, stage I can be replaced by some pragmatic approaches. For example, we may first perform conventional dose escalation to identify the MTD with backfill or followed by small dose expansion on the MTD and a few doses below the MTD to identify the RP2S for stage II randomization. This approach is less efficient to identify the RP2S, but might be more accessible to some practitioners as it is more similar to the conventional paradigm.

\begin{figure}
\caption{Dose-response curves for the nine scenarios in the simulation study. The green, blue, and red lines are the toxicity, PD, and ORR curves, respectively. The circled doses are the optimal doses.}
\begin{center}
\includegraphics[width=1\textwidth]{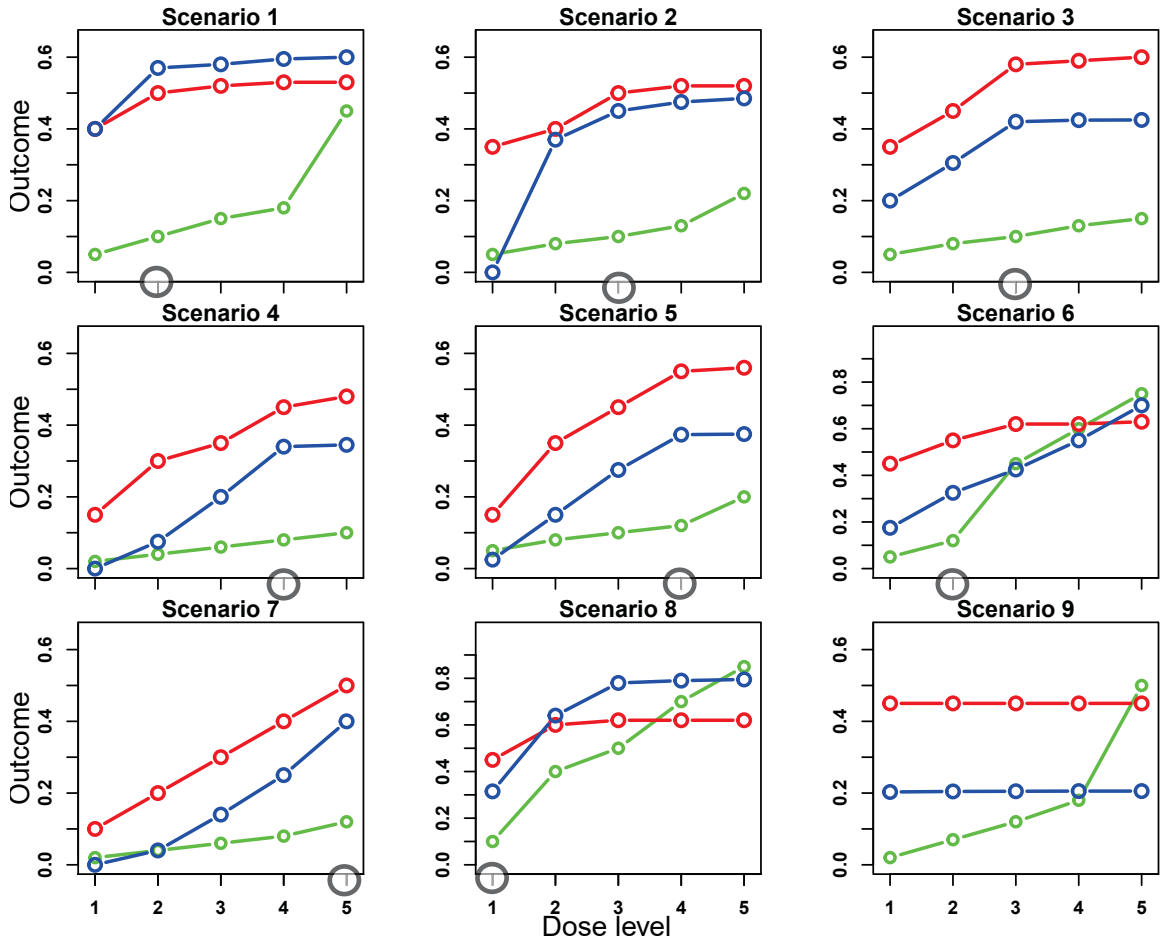}
\end{center}
\end{figure}

\begin{figure}
\caption{Barplots of the selection percentage of each dose under the DROID designs. For each scenario, the four bars represent DROID-BOIN and DROID-CRM using posterior probabilities to select the optimal dose, and DROID-BOIN and DROID-CRM using point estimates to select the optimal dose.}
\begin{center}
\includegraphics[width=1\textwidth]{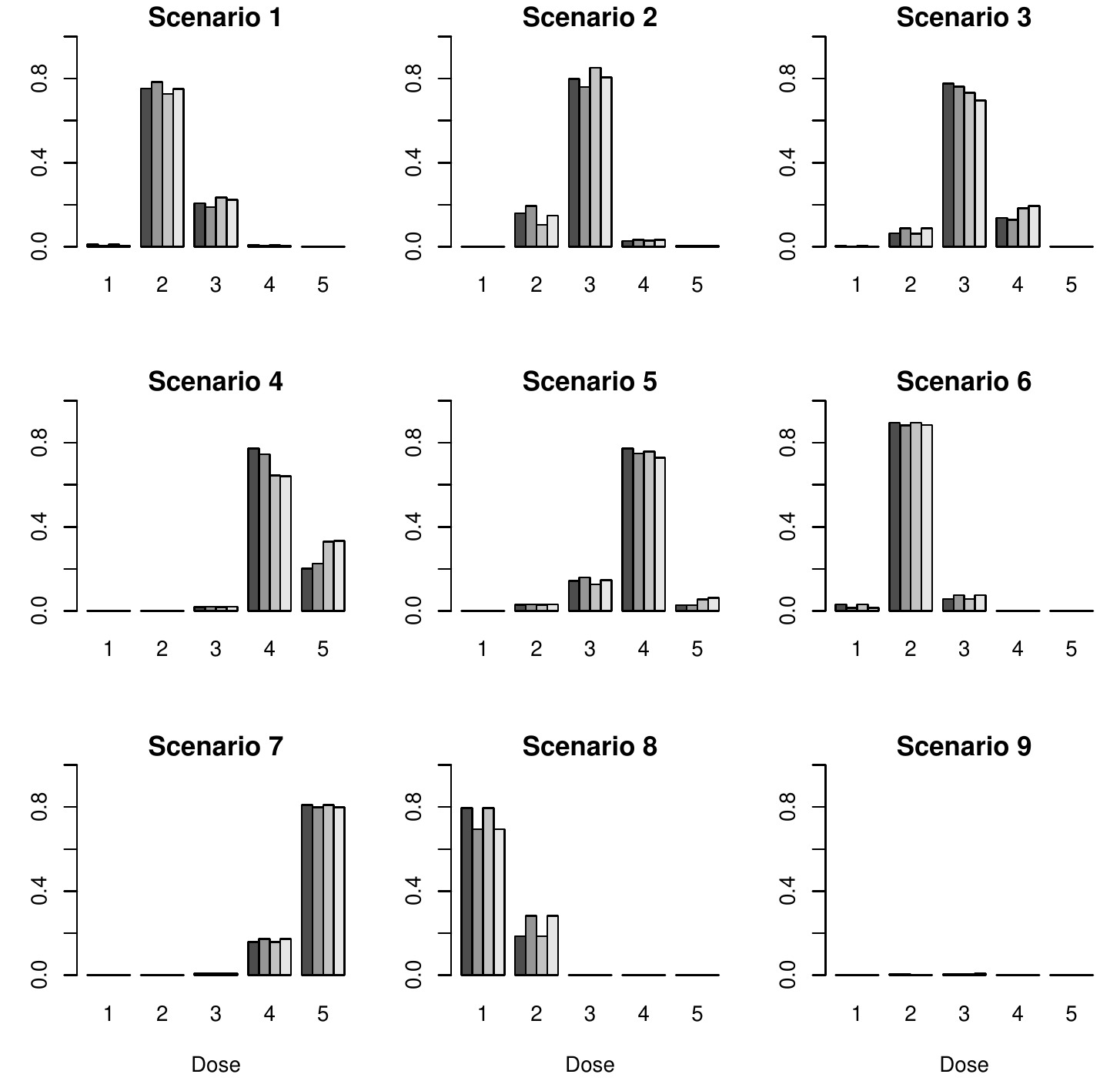}
\end{center}
\end{figure}

\begin{figure}
\caption{Barplots of the selection percentage of each dose under the DROID designs that allow adding doses to TDR/RP2S in stage II. For each scenario, the four bars represent DROID-BOIN and DROID-CRM using posterior probabilities to select the optimal dose, and DROID-BOIN and DROID-CRM using point estimates to select the optimal dose.}
\begin{center}
\includegraphics[width=1\textwidth]{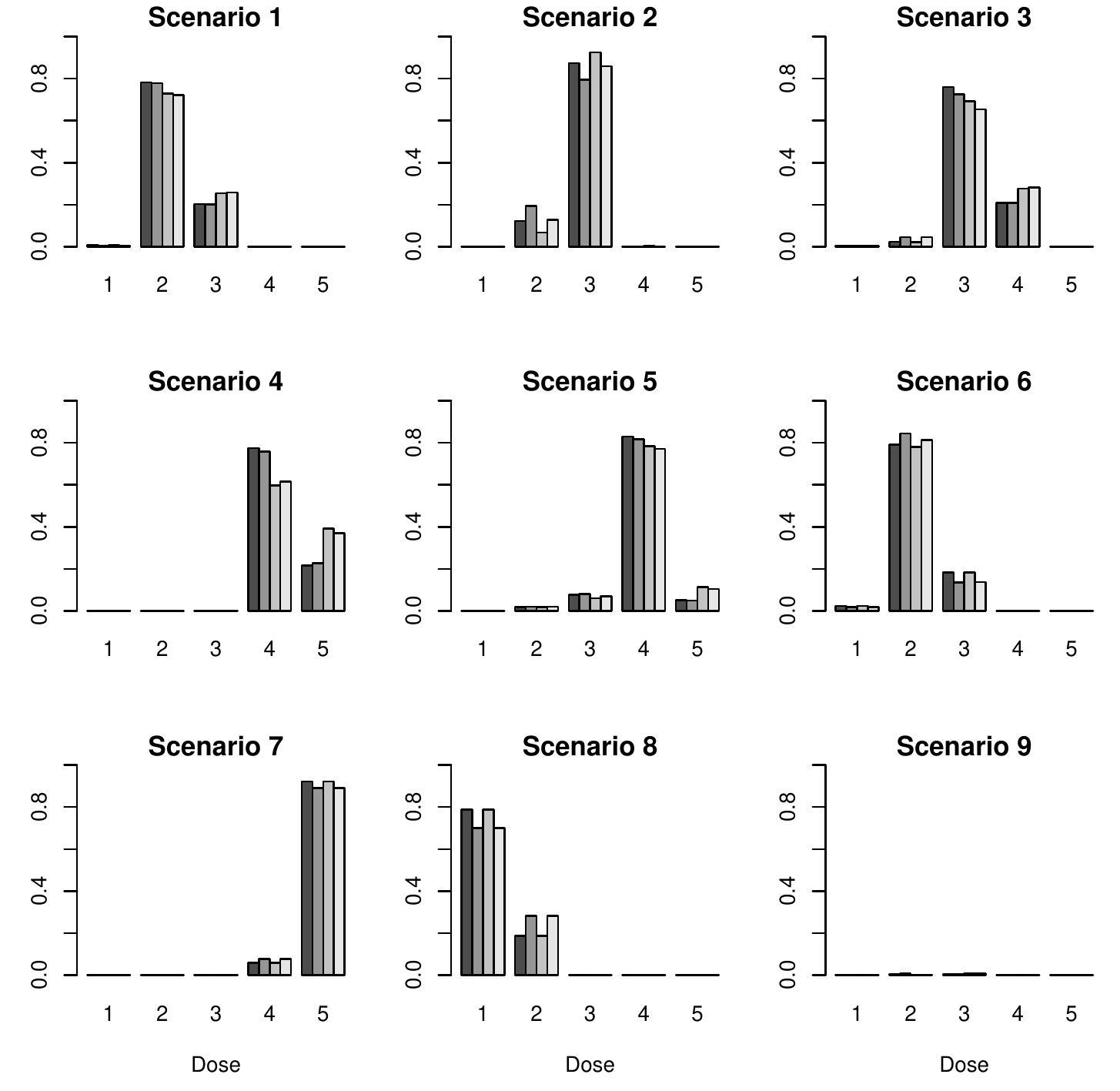}
\end{center}
\end{figure}

\begin{table}
\caption{True toxicity, PD, and ORR at each dose for the nine scenarios. The boldface numbers are the optimal doses.}
\begin{center}
\begin{tabular}{lrrrrrrrrrrr}
\hline
\hline
Dose level &          1 &          2 &          3 &          4 &          5 &            &          1 &          2 &          3 &          4 &          5 \\
\hline
           &            \multicolumn{5}{c}{scenario 1} &&   \multicolumn{5}{c}{scenario 2}     \\

  Toxicity &       0.05 & {\bf 0.10} &       0.15 &       0.18 &       0.45 &            &       0.05 &       0.08 & {\bf 0.10} &       0.13 &       0.22 \\

       ORR &       0.40 & {\bf 0.50} &       0.52 &       0.53 &       0.53 &            &       0.35 &       0.40 & {\bf 0.50} &       0.52 &       0.52 \\

        PD &       0.40 & {\bf 0.57} &       0.58 &       0.60 &       0.60 &            &       0.00 &       0.37 & {\bf 0.45} &       0.48 &       0.49 \\

           &            &     {\bf } &            &            &            &            &            &            &     {\bf } &            &            \\

           &            \multicolumn{5}{c}{scenario 3} &&   \multicolumn{5}{c}{scenario 4}     \\

  Toxicity &       0.05 &       0.08 & {\bf 0.10} &       0.13 &       0.15 &            &       0.02 &       0.04 &       0.06 & {\bf 0.08} &       0.10 \\

       ORR &       0.35 &       0.45 & {\bf 0.58} &       0.59 &       0.60 &            &       0.15 &       0.30 &       0.35 & {\bf 0.45} &       0.48 \\

        PD &       0.20 &       0.31 & {\bf 0.42} &       0.42 &       0.43 &            &       0.00 &       0.08 &       0.20 & {\bf 0.34} &       0.35 \\

           &            &            &     {\bf } &            &            &            &            &            &            &     {\bf } &            \\

           &            \multicolumn{5}{c}{scenario 5} &&   \multicolumn{5}{c}{scenario 6}     \\

  Toxicity &       0.05 &       0.08 &       0.10 & {\bf 0.12} &       0.20 &            &       0.05 & {\bf 0.12} &       0.45 &       0.60 &       0.75 \\

       ORR &       0.15 &       0.35 &       0.45 & {\bf 0.55} &       0.56 &            &       0.45 & {\bf 0.55} &       0.62 &       0.62 &       0.63 \\

        PD &       0.03 &       0.15 &       0.28 & {\bf 0.37} &       0.38 &            &       0.18 & {\bf 0.33} &       0.43 &       0.55 &       0.70 \\

           &            &            &            &     {\bf } &            &            &            &     {\bf } &            &            &            \\

           &            \multicolumn{5}{c}{scenario 7} &&   \multicolumn{5}{c}{scenario 8}     \\

  Toxicity &       0.02 &       0.04 &       0.06 &       0.08 & {\bf 0.12} &            & {\bf 0.10} &       0.40 &       0.50 &       0.70 &       0.85 \\

       ORR &       0.10 &       0.20 &       0.30 &       0.40 & {\bf 0.50} &            & {\bf 0.45} &       0.60 &       0.62 &       0.62 &       0.62 \\

        PD &       0.00 &       0.04 &       0.14 &       0.25 & {\bf 0.40} &            & {\bf 0.32} &       0.64 &       0.78 &       0.79 &       0.80 \\

           &            &            &            &            &     {\bf } &            &     {\bf } &            &            &            &            \\

           &            \multicolumn{5}{c}{scenario 9} && &&&&     \\

  Toxicity &       0.02 &       0.07 &       0.12 &       0.18 &       0.50 &            &            &            &            &            &            \\

       ORR &       0.45 &       0.45 &       0.45 &       0.45 &       0.45 &            &            &            &            &            &            \\

        PD &       0.20 &       0.20 &       0.21 &       0.21 &       0.21 &            &            &            &            &            &            \\
\hline
\hline
\end{tabular}  
\end{center}
\end{table}

\begin{table}
\caption{Selection percentage and the average number of patients treated at each dose under DROID designs using BOIN or CRM in stage I, EffTox and CRM designs. The boldface numbers are optimal doses.}
\begin{small}
\begin{center}
\begin{tabular}{lrrrrrrrrrrr}
\hline
\hline
Dose level &          1 &          2 &          3 &          4 &          5 &            &          1 &          2 &          3 &          4 &          5 \\
\hline
           &            \multicolumn{5}{c}{scenario 1} &&   \multicolumn{5}{c}{scenario 2}     \\

DROID-BOIN Sel \% &      0.012 & {\bf 0.752} &      0.207 &      0.009 &      0.000 &            &      0.000 &      0.160 & {\bf 0.799} &      0.027 &      0.004 \\

no. patients &       20.0 & {\bf 18.4} &       17.6 &       12.3 &        3.5 &            &        6.8 &       19.8 & {\bf 18.9} &       17.0 &        9.1 \\

DROID-CRM Sel \% &      0.005 & {\bf 0.784} &      0.189 &      0.004 &      0.000 &            &      0.000 &      0.195 & {\bf 0.760} &      0.032 &      0.003 \\

no. patients &       20.0 & {\bf 18.8} &       17.4 &       11.4 &        3.0 &            &       16.5 &       18.3 & {\bf 18.2} &       15.6 &        7.6 \\

EffTox Sel \% &      0.180 & {\bf 0.350} &      0.310 &      0.140 &      0.020 &            &      0.170 &      0.140 & {\bf 0.300} &      0.210 &      0.170 \\

no. patients &       16.8 & {\bf 25.4} &       24.2 &       10.6 &        3.8 &            &       16.4 &       13.1 & {\bf 21.8} &       15.2 &       14.2 \\

CRM Sel \% &      0.001 & {\bf 0.051} &      0.168 &      0.719 &      0.061 &            &      0.000 &      0.009 & {\bf 0.045} &      0.306 &      0.640 \\

no. patients &        3.8 &  {\bf 5.1} &        6.4 &       11.3 &        9.4 &            &        3.7 &        4.3 &  {\bf 4.7} &        6.1 &       17.1 \\

           &            &     {\bf } &            &            &            &            &            &            &     {\bf } &            &            \\

           &            \multicolumn{5}{c}{scenario 3} &&   \multicolumn{5}{c}{scenario 4}     \\

DROID-BOIN Sel \% &      0.004 &      0.064 & {\bf 0.777} &      0.138 &      0.002 &            &      0.000 &      0.000 &      0.018 & {\bf 0.773} &      0.201 \\

no. patients &       19.7 &       18.0 & {\bf 18.7} &       16.4 &        9.7 &            &        4.1 &       11.5 &       18.3 & {\bf 18.9} &       17.2 \\

DROID-CRM Sel \% &      0.001 &      0.087 & {\bf 0.762} &      0.129 &      0.001 &            &      0.000 &      0.000 &      0.021 & {\bf 0.746} &      0.226 \\

no. patients &       19.7 &       17.8 & {\bf 18.4} &       15.4 &        8.0 &            &        5.2 &       12.9 &       17.8 & {\bf 18.8} &       16.5 \\

EffTox Sel \% &      0.090 &      0.110 & {\bf 0.280} &      0.220 &      0.300 &            &      0.000 &      0.050 &      0.130 & {\bf 0.240} &      0.580 \\

no. patients &       11.0 &       11.9 & {\bf 20.4} &       15.7 &       21.7 &            &        4.7 &        6.2 &       12.4 & {\bf 16.3} &       41.0 \\

CRM Sel \% &      0.000 &      0.011 & {\bf 0.045} &      0.186 &      0.758 &            &      0.000 &      0.001 &      0.005 & {\bf 0.038} &      0.956 \\

no. patients &        3.7 &        4.3 &  {\bf 4.7} &        5.6 &       17.7 &            &        3.2 &        3.5 &        3.5 &  {\bf 4.1} &       21.7 \\

           &            &            &     {\bf } &            &            &            &            &            &            &     {\bf } &            \\

           &            \multicolumn{5}{c}{scenario 5} &&   \multicolumn{5}{c}{scenario 6}     \\

DROID-BOIN Sel \% &      0.000 &      0.029 &      0.142 & {\bf 0.772} &      0.028 &            &      0.032 & {\bf 0.895} &      0.057 &      0.000 &      0.000 \\

no. patients &        7.3 &       17.7 &       18.0 & {\bf 17.3} &        9.8 &            &       20.0 & {\bf 19.6} &        8.8 &        1.4 &        0.1 \\

DROID-CRM Sel \% &      0.000 &      0.031 &      0.160 & {\bf 0.749} &      0.028 &            &      0.016 & {\bf 0.882} &      0.075 &      0.001 &      0.000 \\

no. patients &        6.7 &       17.5 &       17.9 & {\bf 16.7} &        8.3 &            &       20.1 & {\bf 19.5} &       10.4 &        2.1 &        0.2 \\

EffTox Sel \% &      0.000 &      0.060 &      0.200 & {\bf 0.340} &      0.390 &            &      0.250 & {\bf 0.720} &      0.020 &      0.000 &      0.000 \\

no. patients &        4.2 &        8.0 &       17.1 & {\bf 22.5} &       28.9 &            &       21.4 & {\bf 50.1} &        8.0 &        0.9 &        0.2 \\

CRM Sel \% &      0.000 &      0.012 &      0.039 & {\bf 0.265} &      0.684 &            &      0.013 & {\bf 0.846} &      0.139 &      0.002 &      0.000 \\

no. patients &        3.7 &        4.3 &        4.7 &  {\bf 5.7} &       17.5 &            &        4.0 & {\bf 13.3} &       15.8 &        2.7 &        0.2 \\
           &            &            &            &            &     {\bf } &            &     {\bf } &            &            &            &            \\

           &            \multicolumn{5}{c}{scenario 7} &&   \multicolumn{5}{c}{scenario 8}     \\

DROID-BOIN Sel \% &      0.000 &      0.000 &      0.007 &      0.157 & {\bf 0.810} &            & {\bf 0.796} &      0.185 &      0.000 &      0.000 &      0.000 \\

no. patients &        3.7 &        7.8 &       15.7 &       18.1 & {\bf 17.4} &            & {\bf 23.7} &       12.6 &        2.7 &        0.3 &        0.0 \\

DROID-CRM Sel \% &      0.000 &      0.000 &      0.009 &      0.173 & {\bf 0.798} &            & {\bf 0.694} &      0.283 &      0.000 &      0.000 &      0.000 \\

no. patients &        5.2 &        7.2 &       15.4 &       17.9 & {\bf 16.8} &            & {\bf 21.9} &       15.2 &        3.2 &        0.4 &        0.0 \\

EffTox Sel \% &      0.000 &      0.010 &      0.090 &      0.190 & {\bf 0.700} &            & {\bf 0.710} &      0.190 &      0.000 &      0.000 &      0.000 \\

no. patients &        4.1 &        4.4 &        9.9 &       14.6 & {\bf 47.2} &            & {\bf 45.7} &       24.7 &        3.2 &        0.5 &        0.1 \\

CRM Sel \% &      0.000 &      0.001 &      0.005 &      0.051 & {\bf 0.943} &            & {\bf 0.804} &      0.186 &      0.003 &      0.006 &      0.000 \\

no. patients &        3.3 &        3.5 &        3.6 &        4.1 & {\bf 21.5} &            & {\bf 11.9} &       20.4 &        3.3 &        0.5 &        0.0 \\

           &            &            &            &            &     {\bf } &            &     {\bf } &            &            &            &            \\

           &            \multicolumn{5}{c}{scenario 9} &&&&&&     \\

DROID-BOIN Sel \% &      0.000 &      0.004 &      0.004 &      0.000 &      0.000 &            &            &            &            &            &            \\

no. patients &       19.7 &       18.0 &       18.7 &       15.0 &        3.6 &            &            &            &            &            &            \\

DROID-CRM Sel \% &      0.001 &      0.004 &      0.004 &      0.000 &      0.000 &            &            &            &            &            &            \\

no. patients &       19.8 &       18.3 &       18.9 &       15.9 &        4.7 &            &            &            &            &            &            \\

EffTox Sel \% &      0.340 &      0.260 &      0.310 &      0.080 &      0.010 &            &            &            &            &            &            \\

no. patients &       26.9 &       20.3 &       23.6 &        8.1 &        2.2 &            &            &            &            &            &            \\

CRM Sel \% &      0.000 &      0.011 &      0.116 &      0.829 &      0.044 &            &            &            &            &            &            \\

no. patients &        3.3 &        4.0 &        5.0 &       13.7 &       10.1 &            &            &            &            &            &            \\

\hline
\hline
\end{tabular}  
\end{center}
\end{small}
\end{table}

\begin{table}
\caption{Selection percentage and the average number of patients treated at each dose under DROID designs that allow adding dose to TDR/RP2S in stage II. The boldface numbers are optimal doses.}
\begin{small}
\begin{center}
\begin{tabular}{lrrrrrrrrrrr}
\hline
\hline
Dose level &          1 &          2 &          3 &          4 &          5 &            &          1 &          2 &          3 &          4 &          5 \\
\hline
           &            \multicolumn{5}{c}{scenario 1} &&   \multicolumn{5}{c}{scenario 2}     \\

DROID-BOIN Sel \% &      0.009 & {\bf 0.781} &      0.203 &      0.000 &      0.000 &            &      0.000 &      0.122 & {\bf 0.873} &      0.001 &      0.000 \\

no. patients &       20.2 & {\bf 19.9} &       18.9 &       15.5 &        4.3 &            &        6.8 &       20.0 & {\bf 19.8} &       18.9 &       12.3 \\

DROID-CRM Sel \% &      0.006 & {\bf 0.778} &      0.202 &      0.000 &      0.000 &            &      0.000 &      0.194 & {\bf 0.794} &      0.000 &      0.000 \\

no. patients &       20.2 & {\bf 19.8} &       18.5 &       14.8 &        3.7 &            &       16.5 &       19.9 & {\bf 19.1} &       17.6 &        9.8 \\

           &            &     {\bf } &            &            &            &            &            &            &     {\bf } &            &            \\

           &            \multicolumn{5}{c}{scenario 3} &&   \multicolumn{5}{c}{scenario 4}     \\

DROID-BOIN Sel \% &      0.003 &      0.024 & {\bf 0.759} &      0.209 &      0.000 &            &      0.000 &      0.000 &      0.001 & {\bf 0.774} &      0.216 \\

no. patients &       20.1 &       19.9 & {\bf 19.7} &       18.3 &       11.5 &            &        4.2 &       11.8 &       19.9 & {\bf 19.8} &       19.0 \\

DROID-CRM Sel \% &      0.004 &      0.046 & {\bf 0.724} &      0.209 &      0.000 &            &      0.000 &      0.000 &      0.002 & {\bf 0.759} &      0.227 \\

no. patients &       20.2 &       19.9 & {\bf 19.1} &       17.4 &       10.0 &            &        5.3 &       12.9 &       19.8 & {\bf 19.8} &       18.5 \\

           &            &            &     {\bf } &            &            &            &            &            &            &     {\bf } &            \\

           &            \multicolumn{5}{c}{scenario 5} &&   \multicolumn{5}{c}{scenario 6}     \\

DROID-BOIN Sel \% &      0.000 &      0.018 &      0.078 & {\bf 0.829} &      0.051 &            &      0.024 & {\bf 0.791} &      0.183 &      0.000 &      0.000 \\

no. patients &        7.6 &       19.3 &       19.5 & {\bf 18.6} &       12.4 &            &       20.2 & {\bf 19.7} &       11.9 &        1.5 &        0.0 \\

DROID-CRM Sel \% &      0.000 &      0.021 &      0.081 & {\bf 0.816} &      0.050 &            &      0.019 & {\bf 0.844} &      0.136 &      0.000 &      0.000 \\

no. patients &        6.4 &       19.5 &       19.4 & {\bf 18.1} &       10.7 &            &       20.1 & {\bf 19.4} &       12.1 &        2.1 &        0.2 \\

           &            &            &            &     {\bf } &            &            &            &     {\bf } &            &            &            \\

           &            \multicolumn{5}{c}{scenario 7} &&   \multicolumn{5}{c}{scenario 8}     \\

DROID-BOIN Sel \% &      0.000 &      0.000 &      0.002 &      0.059 & {\bf 0.921} &            & {\bf 0.788} &      0.187 &      0.001 &      0.000 &      0.000 \\

no. patients &        3.8 &        7.6 &       17.8 &       19.7 & {\bf 18.9} &            & {\bf 23.9} &       12.6 &        2.6 &        0.2 &        0.0 \\

DROID-CRM Sel \% &      0.000 &      0.000 &      0.003 &      0.078 & {\bf 0.890} &            & {\bf 0.701} &      0.283 &      0.000 &      0.000 &      0.000 \\

no. patients &        5.2 &        7.2 &       17.7 &       19.5 & {\bf 18.2} &            & {\bf 21.9} &       15.3 &        3.3 &        0.4 &        0.0 \\

           &            &            &            &            &     {\bf } &            &     {\bf } &            &            &            &            \\

           &            \multicolumn{5}{c}{scenario 9} &&&&&&     \\

DROID-BOIN Sel \% &      0.001 &      0.006 &      0.005 &      0.000 &      0.000 &            &            &            &            &            &            \\

no. patients &       20.1 &       19.8 &       19.7 &       18.2 &        5.5 &            &            &            &            &            &            \\

DROID-CRM Sel \% &      0.000 &      0.007 &      0.004 &      0.000 &      0.000 &            &            &            &            &            &            \\

no. patients &       20.1 &       19.8 &       19.6 &       18.4 &        6.1 &            &            &            &            &            &            \\

\hline
\hline
\end{tabular}  
\end{center}
\end{small}
\end{table}

\label{lastpage}

\end{document}